\documentclass[twocolumn]{aastex631}

\usepackage{natbib}

\usepackage[T1]{fontenc} 

\usepackage{array}

\usepackage{multirow}

\usepackage{graphicx}

\usepackage{booktabs}

\usepackage{placeins}  

\usepackage{hyperref} 

\setcitestyle{authoryear}

\begin{document}

\title{Saturns, but not super-Jupiters, occur more frequently in the presence of inner super-Earths}

\author{Etienne Lefèvre-Forján}
\affiliation{Facultad de Física, Pontificia Universidad Católica de Chile, Av. Vicuña Mackenna 4860, Macul, Santiago, Chile}
\affiliation{UFR Physique, Université Paris Cité, 4 Rue Elsa Morante, F-75013 Paris, France}

\author[0000-0002-1078-9493]{Gijs D. Mulders}
\affiliation{Facultad de Física, Pontificia Universidad Católica de Chile, Av. Vicuña Mackenna 4860, Macul, Santiago, Chile}

\begin{abstract}
    Studies from recent years have reached different conclusions regarding how frequently super-Earths are accompanied by long period giant planets and vice versa. This relation has been predicted to be mass dependent by planet formation models. We investigate that as the origin of the discrepancy using a radial velocity sample: the California Legacy Survey. We perform detection completeness corrections in order to discard detection bias as a possible explanation to our results. After bias corrections, we find that cold Jupiters are $5.65^{+1.08}_{-2.57}$ times more massive when not in company of an inner super-Earth, while super-Earths are not significantly more massive while in company of an outer giant planet. We also report an occurrence enhancement for Saturns (median projected mass of $0.6M_J$) while in presence of a super-Earth by a factor of ${\sim}4$, and for super-Earths in presence of Saturns by the same factor. This positive correlation disappears for super-Jupiters (median projected mass of $3.1M_J$). These results show that while cold Jupiters are generally accompanied by inner super-Earths, this does not hold for the largest giant planets, such as those that will be discovered by Gaia, which will likely not be accompanied by transiting planets. The mass dependence, in combination with different detection limits of different surveys, may explain the discrepancies concerning occurrence relations between cold Jupiters and super-Earths.
\end{abstract}

\section{Introduction}
\label{sec:1}

Interactions between hot super-Earths and cold giants have an impact on the formation and evolution of planetary systems. A positive correlation between those types of planets contradicts what can be seen in the Solar System, where super-Earths are absent. Studies on the relation between super-Earths and cold giants have shown quite different conclusions, some of them reporting a strong positive correlation \citep[e.g.][]{zhu2018, bryan2019} and others finding neutral or negative correlations \citep[e.g.][]{bonomo}. Knowing the sign of that correlation would help us understand how these planets form and interact, so different variables have been proposed as possible causes to the discrepancy, such as metallicity \citep[e.g.][]{bryan2024, gajendran} or gap complexity \citep[e.g.][]{he2023}.

The first calculations for the correlation between outer gas giants and inner super-Earths were done by \citet{zhu2018} and \citet{bryan2019}, who agreed on a strong correlation. \citet{zhu2018} reported that $90^{+20}_{-20}\%$ of stars with Jupiter analogs also host an inner super-Earth, and \citet{bryan2019} reported $102^{+34}_{-52}\%$ for that same measure. After that, \citet{rosenthal2022} showed a weaker but still positive correlation, reporting that $42^{+17}_{-13}\%$ of cold giant hosts also host an inner small planet. On the other hand, a negative correlation was found by \citet{bonomo}, as they reported an occurrence rate for cold Jupiters with inner small companions of $9.3^{+7.7}_{-2.9}\%$, which is lower than the $20.2^{+6.3}_{-3.4}\%$ frequency obtained for all cold Jupiters from the studied sample \citep{wittenmyer2020}. Given the range of the results. we study the correlation in search of explanations for this discrepancy.

Theoretical models also make different predictions: \citet{schlecker} anticipated that super-Earths are more likely to be destroyed while in company of outer giants; \citet{chachan2022} showed that outer giants and super-earths could be formed simultaneously; \citet{bitsch} proposed that cold Jupiters could reduce the survival rate of super-Earths; and \citet{best} reported that solid material can be transported to the inner system in the presence of a cold Jupiter, allowing formation of short-period planets. These models provide a variety of possible explanations to the observed interactions between super-Earths and cold Jupiters.

As a possible solution to the observational discrepancies, \citet{mulders} proposed a mass dependence: in systems with both super-Earths and gas giants, the planet's masses should be anti-correlated (i.e. the most massive super-Earths should not exist in systems that also harbor the most massive giant planets). That was accomplished by simulating the growth of planet cores using a pebble accretion model \citep[][]{ormel, lambrechts2012}, showing that gas giants tend to block the flow of pebbles, so the super-Earths can't grow that much. This proposal leads us to the question we will try to answer in this paper: Does the presence of an outer giant companion affect the mass of small inner planets? And also: Does the presence of an inner small companion affect the mass of outer giant planets?

We use the California Legacy Survey (CLS) sample \citep{rosenthal2021} to compare the mass of super-Earths and cold Jupiters given the presence (or absence) of their counterparts. In \hyperref[sec:2]{section 2} we present the sample, and compare the projected masses of super-Earths with and without cold Jupiter companions (and vice versa). In \hyperref[sec:3]{section 3} we perform detection completeness corrections for each sample (super-Earths with and without cold Jupiter companions, and cold Jupiters with and without super-Earth companions) in order to account for detection bias. In \hyperref[sec:4]{section 4} we calculate occurrence rates using the intrinsic numbers of planets calculated during the detection completeness corrections, for the purpose of testing the correlation's dependence on giant planet mass. In \hyperref[sec:5]{section 5} we propose mass dependence as a solution to the discrepancies on the correlation between super-Earths and cold Jupiters, and explain how our results fit in the overall picture of exoplanet demographics and planetary systems' architectures.

\section{Data sample}
\label{sec:2}

We used the sample presented in the California Legacy Survey Catalog from \citet{rosenthal2021} to compare the projected masses of super-Earths (SE) with and without outer giant companions, and cold Jupiters (CJ) with and without inner super-Earth companions. 719 FGKM type stars were selected in order to get a sample mostly unbiased for known exoplanets. The targets were monitored via the radial velocity (RV) method during three decades, providing physical and orbital data for 178 exoplanets from hot super-Earths (${\sim}10M_\oplus$) to cold gas giants.
They also characterized the detection completeness of each target's dataset, and of the entire survey, by running injection-recovery tests \citep{fulton}. This will be useful for the calculations in \hyperlink{sec:3}{section 3}. In the following sub-sections, we will investigate if the masses of cold Jupiters and super-Earths are affected when in company of their counterparts.

\begin{figure}
\label{fig:1}
    \centering
    \includegraphics[width=240pt]{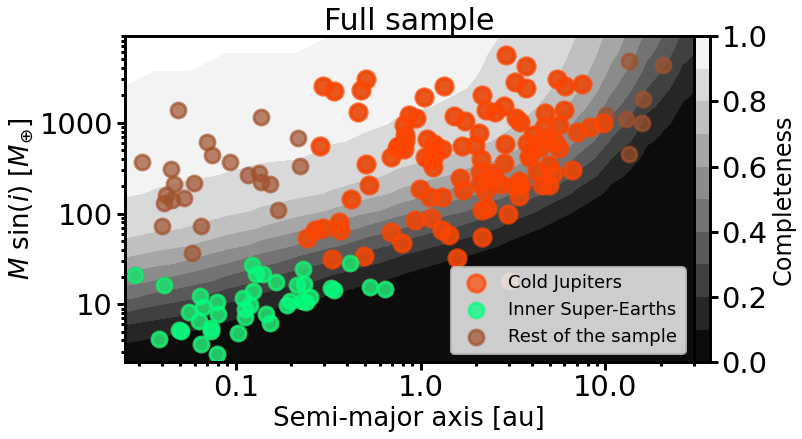}
    \caption{Exoplanet sample from the California Legacy Survey. Green and red circles show super-Earths and cold Jupiters respectively, as defined in \hyperref[sec:2.1]{section 2.1}. Contours represent detection completeness.}
\end{figure}

\subsection{Super-Earth mass when an outer giant is present}
\label{sec:2.1}

In order to study super-Earths and cold Jupiters, we define them the same way as \cite{rosenthal2022}, as it is based on the same data: cold Jupiters are planets with a projected mass within the range $30\mbox{-}6000 M_\oplus$ and super-Earths are planets with a projected mass lower than $30 M_\oplus$. For the analysis of super-Earths, we put an average distance to host boundary of $1$ au, and consider as an outer giant companion any giant planet with an average distance to host higher than the super-Earth.

\begin{figure}
\label{fig:2}
    \centering
    \includegraphics[width=242pt]{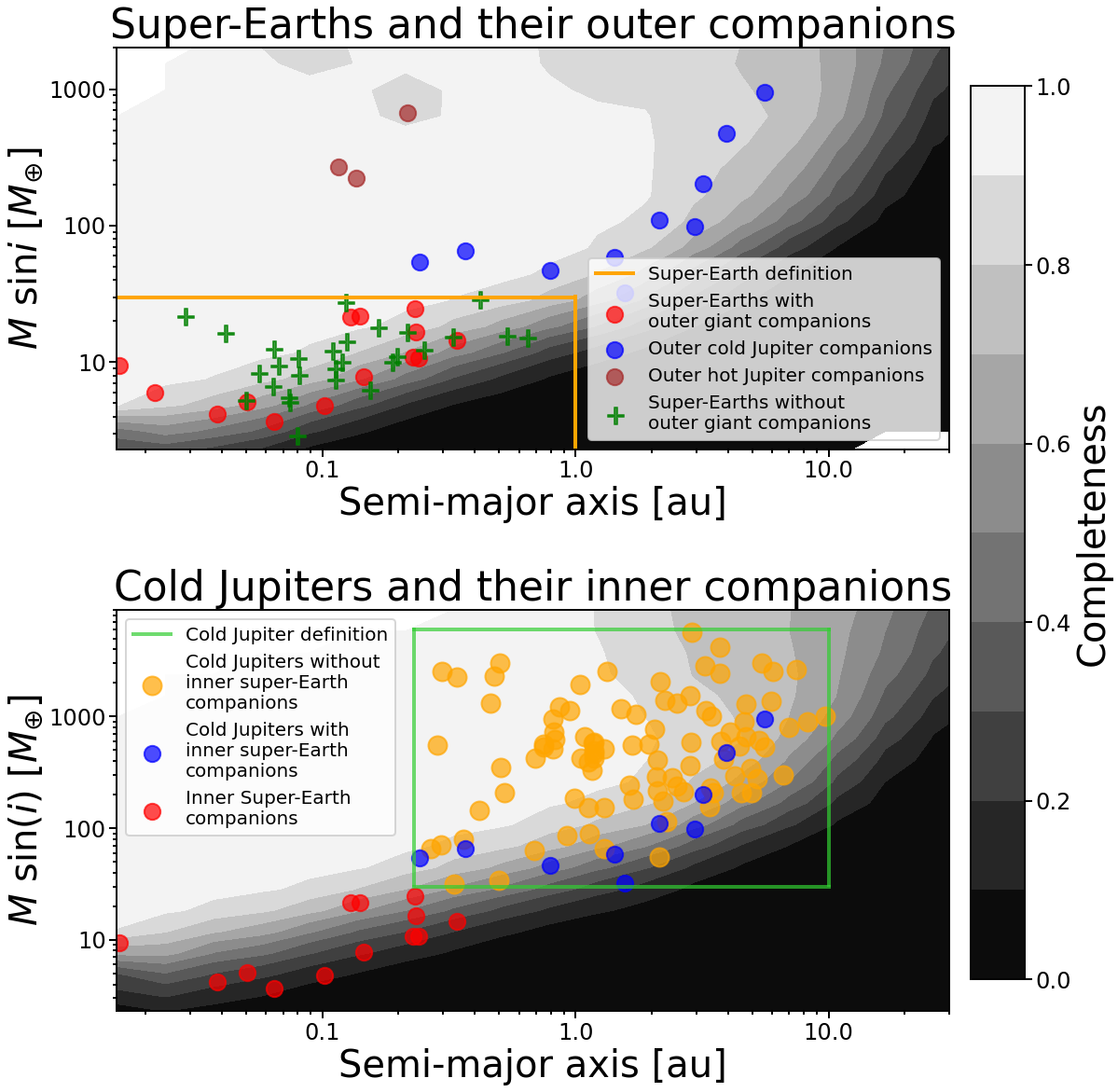}
    \caption{Super-Earths and their outer giant companions (top) \& cold Jupiters and their inner super-Earth companions (bottom). Completeness contours here are obtained from systems that host at least one super-Earth (top) and systems that host at least one cold Jupiter (bottom), and differ from those in \hyperref[fig:1]{figure 1}. Based on figure 1 from \citet{rosenthal2022}, data taken from \citet{rosenthal2021}.}
\end{figure}

We split the super-Earth sample into two parts:
super-Earths with outer giant companions and super-Earths without outer giant companions, as shown in \hyperref[fig:2]{figure 2}. This allows us to compare the projected masses of the super-Earths in both of these sub-samples. Notice that some of these super-Earths may share a host, so there might be fewer hosts than planets for this sample. Indeed, after counting the number of stars hosting these 42 super-Earths we get a total of 29 hosts, which will be relevant for the calculation of occurrence rates.

\begin{figure}
\label{fig:3}
    \centering    \includegraphics[width=240pt]{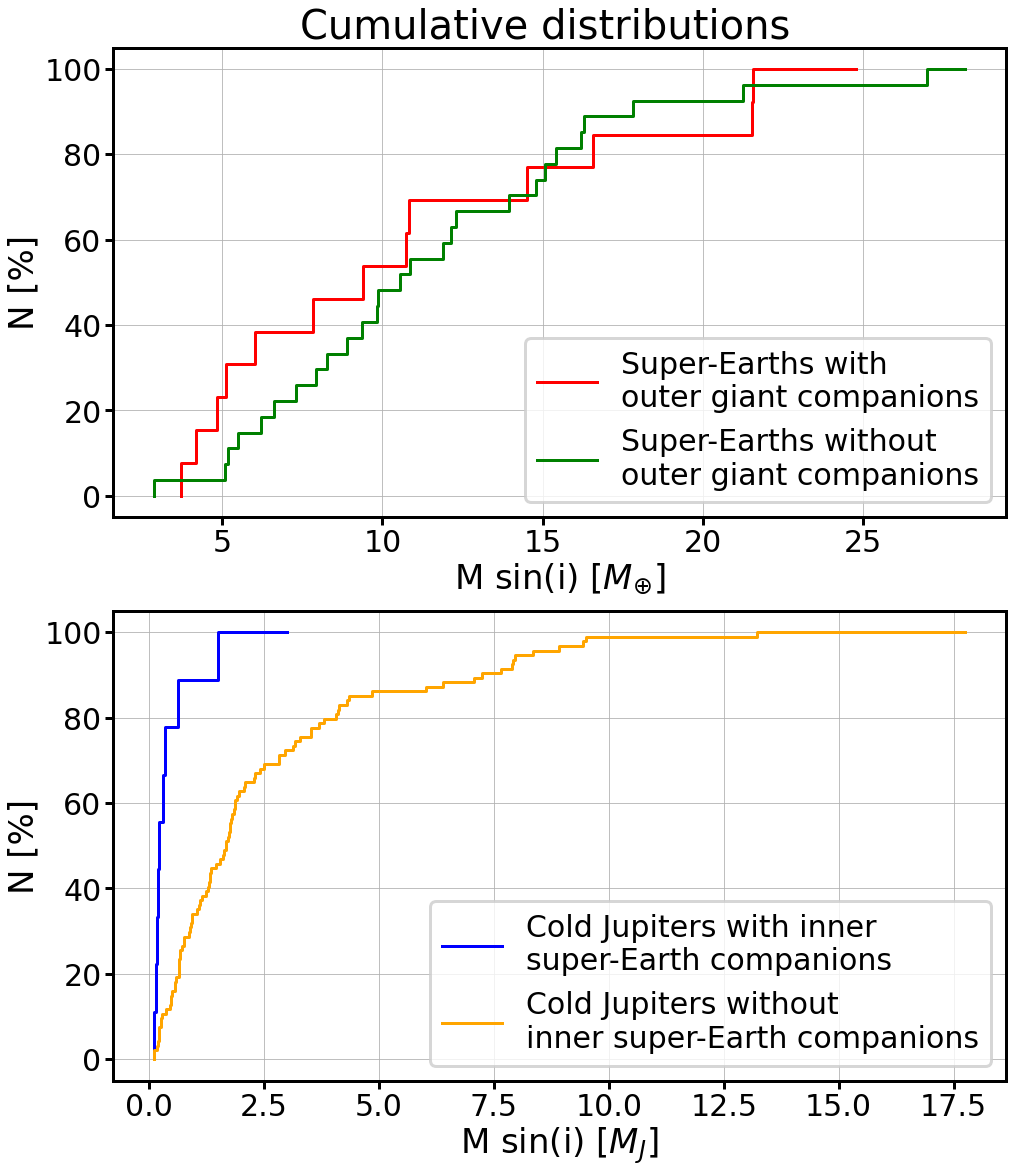}    \caption{Cumulative distributions for each sub-sample of super-Earths (up) and for each sub-sample of cold Jupiters (down). In the super-Earths' case, either the distributions nor the mass threshold is much different from one sub-sample to another. In the cold Jupiters case, the ones with super-Earth companions seem to have a much lower mass than the ones without super-Earth companions.}
\end{figure}

Next, we calculate the median planet projected mass for each sub-sample, with the purpose of investigating if there is a considerable mass difference for super-Earths with and without outer giant companions. We choose the median because it is less influenced by extremely large values, which would be an important disadvantage for the mean since we are working with data from skewed distributions. The resulting median projected mass for super-Earths with outer giant companions is $10.07^{+2.60}_{-2.36} M_\oplus$ (the error corresponds to the difference between the median and the percentiles 16 and 84 of the bootstrap distribution of the median of the sample, for 10000 iterations), and the median projected mass for super-Earths without outer giant companions is $10.71^{+1.43}_{-0.88} M_\oplus$ (see \hyperref[table:1]{table 1}), so there is no significant difference in mass between super-Earths with outer giant companions and super-Earths without outer giant companions. The cumulative distribution function (\hyperref[fig:3]{figure 3}) show that the mass distributions for each sub-sample are very similar, as the lines even intersect at some points. The Kolmogorov-Smirnov test gives a p-value of ${\sim}0.77$, so the null hypothesis that the two samples are drawn from the same underlying distribution can not be rejected at a significant level.

\begin{table*}
\normalsize
\label{table:1}
    \centering
    \begin{tabular}{ |c|c c c c| }
    \hline
    Sample & Median & With companions & Without companions & Ratio \\ [0.2cm]
    \hline
    \multirow{2}{*}{Super-Earths} & Observed & $10.07^{+2.60}_{-2.36}\, M_\oplus$ & $10.71^{+1.43}_{-0.88}\, M_\oplus$ & $1.06^{+0.42}_{-0.34}$ \\ [0.2cm]
    & Debiased & $10.74^{+5.84}_{-2.92}\, M_\oplus$ & $10.87^{+3.08}_{-1.04}\, M_\oplus$ & $1.01^{+0.84}_{-0.37}$ \\ [0.2cm]
    \hline
    \multirow{2}{*}{Cold Jupiters} & Observed & $0.26^{+0.09}_{-0.07}\, M_J$ & $1.68^{+0.10}_{-0.22}\, M_J$ & $6.46^{+2.62}_{-2.59}$ \\ [0.2cm]
    & Debiased & $0.31^{+0.04}_{-0.12}\, M_J$ & $1.75^{+0.11}_{-0.12}\, M_J$ & $5.65^{+1.08}_{-2.57}$ \\ [0.15cm]
    \hline
    \end{tabular}
    \caption{Median projected mass of each sub-sample. The error bars correspond to the difference between the median and the percentiles 16 and 84 of the bootstrap distribution of the median of the sample (for 10000 iterations). The ratio column shows how super-Earth mass is not significantly affected by the presence of an outer giant, but the cold giants are much more massive when not in company of an inner super-Earth.}
\end{table*}

\subsection{Cold Jupiter mass when a super-Earth is present}
\label{sec:2.2}

The average distance to host boundary we put for cold Jupiters is $0.23$ au to $10$ au \citep[also taken from][]{rosenthal2022}. We will consider a super-Earth companion any planet with less than $30M_\oplus$ with an average distance to host lower than the cold Jupiter. As is done for the super-Earths in the previous sub-section, we split the cold Jupiter sample in two: cold Jupiters with inner super-Earth companions and cold Jupiters without inner super-Earth companions.

Observing the lower plot from \hyperref[fig:2]{figure 2} carefully, it can be noticed that there may be a great difference in median projected mass between cold Jupiter with and without super-Earth companions, as cold Jupiters with inner super-Earths (blue circles) seem to be distributed in a lower mass region than cold Jupiters without inner super-Earths (orange circles). We calculate the median projected mass for each sub-sample: cold Jupiters with super-Earth companions have a median projected mass of $0.26^{+0.09}_{-0.07} M_J$ and cold Jupiters without super-Earth companions have a median projected mass of $1.68^{+0.10}_{-0.22} M_J$
(see \hyperref[table:1]{table 1}). The median projected mass is then $6.46^{+2.62}_{-2.59}$ times higher for the cold Jupiters without super-Earth companions, which suggests that giant planets may be influencing the formation or survival rate of inner small planets.

The cumulative distribution functions in \hyperref[fig:3]{figure 3}
confirm that the typical projected mass for the cold Jupiters without inner super-Earth companions is much higher than the one for the accompanied giants. The Kolmogorov-Smirnov test rejects the possibility that the two samples are drawn from the same distribution at a significance level of ${\sim}0.001$ (p-value). That indicates that indeed the observed giant planet projected masses are significantly lower when accompanied by a super-Earth.

However, the difference in giant planet mass may be a result of detection bias. Super-Earths require a higher sensitivity to be detected than giant planets, and not all stars in the survey have the same sensitivity, either due to the observing strategy or the intrinsic noise level of the stars. Therefore, in stars with lower sensitivity, super-Earths may be missed while massive giant planets are still detectable. This may lead to, on average, higher giant planet masses in systems without super-Earths, as observed. This should be accounted for by performing a correction by detection completeness, as we explain in \hyperref[sec:3]{section 3}.

\section{detection completeness corrections}
\label{sec:3}

Exoplanet detection via the RV method is biased towards large planets at short distances from the star, as they are easier to detect. This bias can be corrected for by quantifying the detection completeness \citep[e.g.][]{cumming, howard, mayor2011}.

We adopt the detection completeness as calculated by \citet{rosenthal2021} using injection recovery tests, publicly accessible on GitHub\footnote{\url{https://github.com/leerosenthalj/CLSI/}}. As these are provided per-star basis, we are able to compute the detection completeness for specific samples. To correct cold Jupiters with super-Earth companions, we need to account for the detection completeness of systems that host super-Earths, as we want to know if the sensibility for detecting cold Jupiters in such systems was high or low. Then, the complement of that sample (i.e. detection completeness of stars that do not host super-Earths) will be used to correct cold Jupiters without super-Earth companions. We apply this same logic to correct super-Earths: we use the detection completeness of cold Jupiter hosts to correct super-Earths with outer giant companions and the detection completeness of stars that do not host cold Jupiters to correct super-Earths without outer giant companions. The detection completeness contours shown in \hyperref[fig:2]{figure 2} are the ones of super-Earth hosts (top panel) and cold Jupiter hosts (bottom panel)\footnote{To visualize the detection completeness difference between cold Jupiters with and without inner super-Earth companions, see \hyperref[appA]{appendix A}.}. For instance, a $10 M_\oplus$ planet at $0.1$ au would have a detection probability around $90\%$ in the super-Earth sample (\hyperref[fig:2]{figure 2} top panel) and around $50\%$ in the cold Jupiter sample (\hyperref[fig:2]{figure 2} bottom panel). This justifies our careful treatment of the completeness function, and as we show below, it impacts the final result.

\subsection{Corrections for cold Jupiters}
\label{sec:3.1}

\begin{figure*}
\label{fig:4}
    \centering
    \includegraphics[width=360pt]{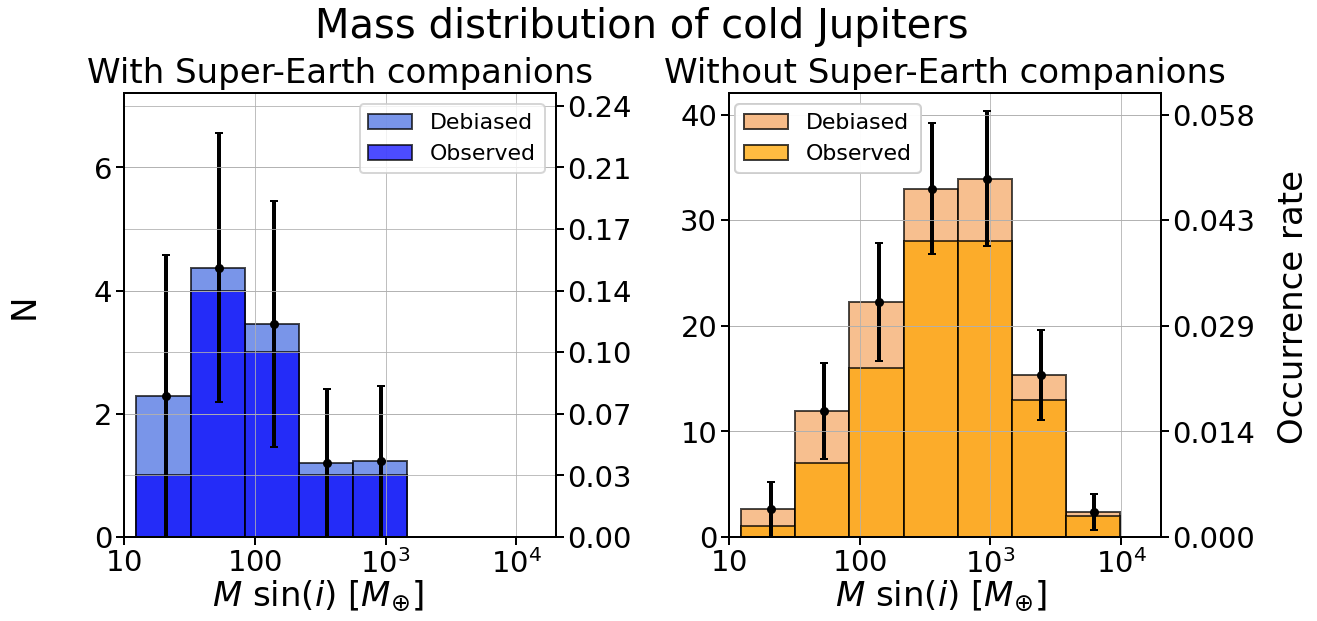}
    \caption{Corrected number of planets for each sub-sample of cold Jupiters. Error bars correspond to the standard error ($1/\sqrt{N}$), and the right vertical axis on both panels shows occurrence rate.}
\end{figure*}
We calculate the average detection completeness per projected mass-semi-major-axis bin, picking $9$ bins evenly spaced in log distance between $1.5 \cdot 10^{-2}$ and $41.6$ au, and $9$ bins evenly spaced in log mass between $1.9$ and $9.8 \cdot 10^3 \, M_\oplus$. Then, we compute an intrinsic number of cold Jupiters by dividing the number of planets by the average completeness inside each bin, obtaining a debiased intrinsic number to work with (for a more detailed explanation of this calculation, see the \hyperref[appA]{appendix A}). This will allow us to determine a corrected value for the median mass.

The results (see \hyperref[table:1]{table 1}) show that the ratio obtained in \hyperref[sec:2.2]{section 2.2} is not a detection bias issue. In fact, the projected mass ratio of cold Jupiters stays close to $6$ after correcting for detection bias. This is a consequence of all the cold Jupiters being located in high completeness regions whether they have companions or not (see \hyperref[fig:2]{figure 2}). Thus, the observed higher mass in cold Jupiters without super-Earth companions is a real feature of the planet population.

\subsection{Corrections for inner super-Earths}
\label{sec:3.2}

\begin{figure*}
\label{fig:5}
    \centering
    \includegraphics[width=360pt]{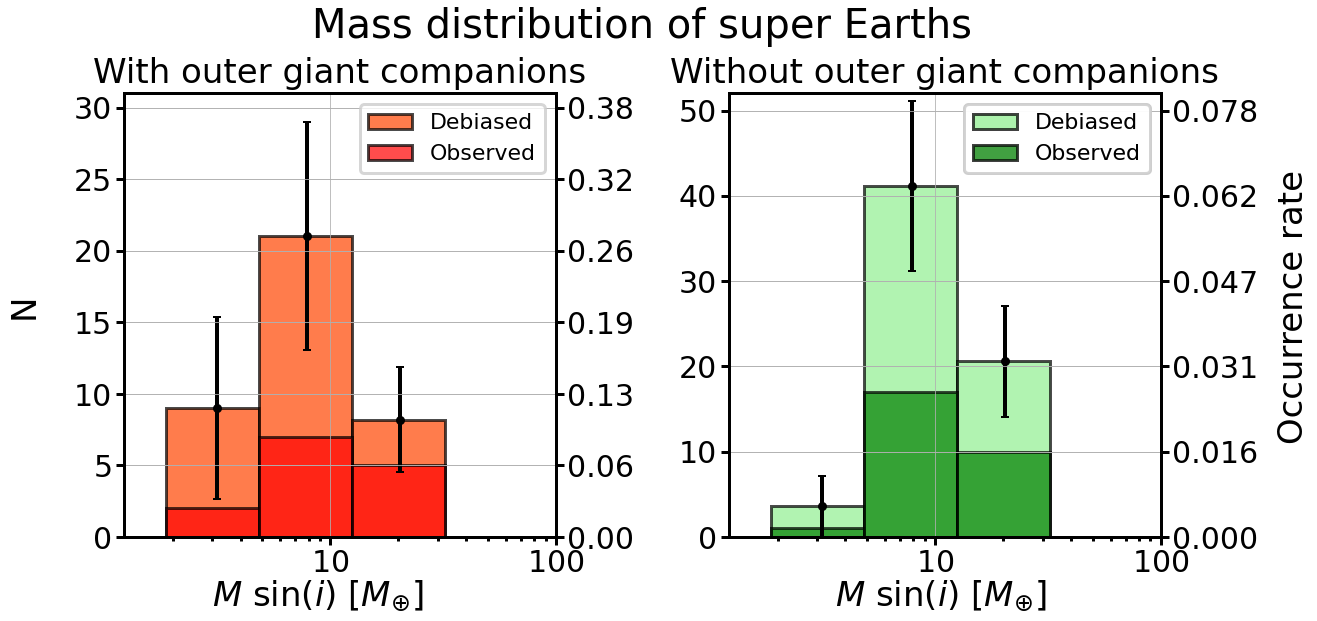}
    \caption{Corrected number of planets for each sub-sample of super-Earths. Error bars correspond to the standard error ($1/\sqrt{N}$), and the right vertical axis on both panels shows occurrence rate.}
\end{figure*}

We also perform the same analysis to the small planet sample, in order to reject the hypothesis that the results from \hyperref[sec:2.1]{section 2.1} are due to a detection bias.

The ratio between median projected masses (shown in \hyperref[table:1]{table 1}) does not change practically, showing that the projected mass of super-Earths with cold Jupiter companions is pretty much the same as the projected mass of super-Earths without cold Jupiter companions. This implies that the mass of the small planets is not strongly affected by detection bias.

\section{Occurrence rates}
\label{sec:4}

We already showed that cold Jupiters without super-Earth companions are significantly more massive than the accompanied cold Jupiters, but here we would like to examine the strength and direction of the correlation. We calculate the occurrence rates to test the correlation between super-Earths and cold Jupiters. In particular, we look for evidences whether the most massive cold Jupiters (super-Jupiters) suppress the formation of super-Earths, or whether the correlation remains but gets weaker.

\begin{figure*}
\label{fig:6}
    \centering
    \includegraphics[width=360pt]{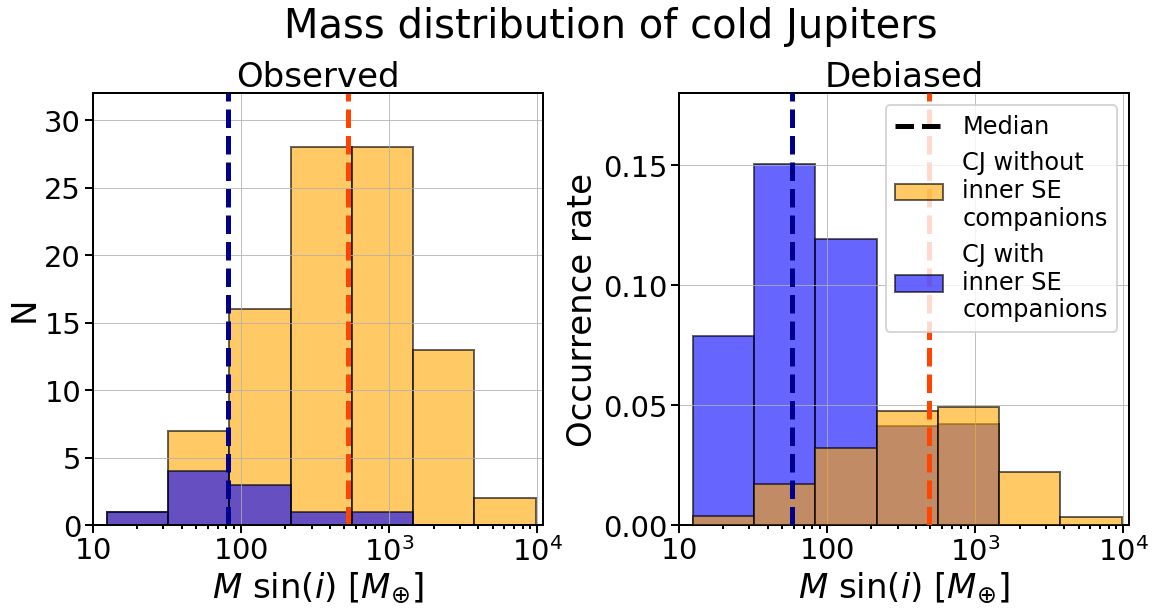}
    \caption{projected mass distribution of cold Jupiters, dashed line representing the median projected mass for each sub-sample. Left panel shows number of planets, right panel shows occurrence rate. It can be noticed that the distance between the dashed lines increases after correction. Right plot shows clearly how the occurrence rate for accompanied cold Jupiters is a lot higher.}
\end{figure*}

In order to get the occurrence rates, we divide the number of planets by the number of hosts, for each type of planet. We do this for the samples after correction to account for detection bias. \hyperref[table:2]{Table 2} (first row) shows how the occurrence rate of cold Jupiters increases from $18.6 \pm 1.8\%$ to $43.2 \pm 13.7\%$ when in company of inner super-Earths. This is complemented by \hyperref[fig:6]{figure 6}, where the difference of occurrence rate between cold Jupiters with and without inner super-Earth companions can be visualized, as well as the difference in projected mass between those groups.

\begin{table*}
\label{table:2}
\small
    \centering
    \begin{tabular}{ |c c c c c c| }
    \hline
    Sample & $P(CJ)$ & $P(CJ|SE)$ & $P(SE|CJ)$ & $P(CJ|SE)/P(CJ)$ & Significance \\ [0.3cm]\hline
    Cold Jupiters & $18.6\pm1.8\%$ & $43.2\pm13.7\%$ & $49.0\pm13.1\%$ & $2.3\pm0.8$ & $1.8 \sigma$ \\ [0.3cm]
    Super-Jupiters & $8.8\pm1.2\%$ & $4.2\pm4.2\%$ & $8.4\pm5.8\%$ & $0.5\pm0.5$ & $1\sigma$ \\ [0.3cm]
    Saturns & $9.8\pm1.4\%$ & $39.0\pm13.0\%$ & $56.6\pm17.1\%$ & $4.0\pm1.4$ & $2.2\sigma$\\
    \hline
    \end{tabular}
\caption{Calculated occurrence rates. Notice that it is much more likely to find a cold Jupiter in a system with a super-Earth than finding a cold Jupiter in the whole sample. This also works when trying to find super-Earths: it is more likely to find one in systems that host at least one cold Jupiter (considering that we got $P(SE) \approx 14.4\pm2.2\%$). This correlation stands for the bottom 50\% least massive cold Jupiters (Saturns; $M<1.5M_J$), but it does not for the top 50\% most massive (Super-Jupiters; $M\geq1.5M_J$). Intrinsic numbers of planets were used, error bars correspond to the standard error.}
\end{table*}

The occurrence rate of super-Earths increases from $14.4 \pm 2.2\%$ to $49.0 \pm 13.1\%$ when in presence of an outer giant companion, showing that super-Earths are enhanced by them. This contradicts with some of the theory predictions outlined in \hyperref[sec:1]{section 1}. To further understand this issue, we test whether only the least massive cold Jupiters are contributing to the observed correlations, while the most massive cold Jupiters are anti-correlated with super-Earths.

To do it, we calculate the occurrence rates of the $50\%$ most massive cold Jupiters (i.e. super-Jupiters, with a median projected mass of $3.1 M_J$) and the $50\%$ least massive cold Jupiters (i.e. Saturns, with a median projected mass of $0.6 M_J$). Results are displayed in \hyperref[table:2]{table 2}. We calculate the enhancement factor by dividing the occurrence rate of cold Jupiters with super-Earth companions ($P(CJ|SE)$) by the occurrence rate of all cold Jupiters ($P(CJ)$). The enhancement factor is ${\sim}4$ for the Saturns and ${\sim}0.5$ for the super-Jupiters. For the super-Earths, we divide the occurrence rate of super-Earths with outer giant companions ($P(SE|CJ)$) by the occurrence rate of all super-Earths ($P(SE)$). The enhancement factor is ${\sim}4$ for the Saturns and ${\sim}0.6$ for the super-Jupiters (we use $P(SE) \approx 14.4 \pm 2.2 \%$, as the occurrence rate of super-Earths does not change for the Saturns and super-Jupiters samples). This shows that the positive correlation disappears when we select the most massive cold Jupiters, but remains when we select the least massive ones.

\section{Conclusions \& discussion}
\label{sec:5}

We have analyzed the California Legacy Survey sample to look for a correlation between the projected mass of inner super-Earths and the presence of outer giant planets, and also between the projected mass of cold Jupiters and the presence of inner super-Earths, taking into account detection completeness. Our main results are:

\begin{itemize}
    \item Cold Jupiters are $5.65^{+1.08}_{-2.57}$ times more massive when not in company of an inner super-Earth, which is a significant mass difference between the two groups.
    \item Super-Earths are $1.01^{+0.84}_{-0.37}$ times more massive when not in company of an outer giant planet, which is not a significant mass difference between the two groups.
    \item Super-Jupiters are not enhanced in occurrence when inner super-Earths are present, while Saturn-mass planets are enhanced by a factor of ${\sim}4$.
    \item Occurrence of super-Earths is enhanced by a factor of ${\sim}4$ when in presence of a Saturn-mass giant, but they are not enhanced when in company of a super-Jupiter.
\end{itemize}

These results suggest that the positive correlation between super-Earths and cold Jupiters only holds for the least massive giants. As the occurrence of super-Earths is only enhanced while in presence of a Saturn-mass giant (and not a super-Jupiter), our results agree with the hypothesis proposed by \citet{mulders}, that the most massive super-Earths should not exist in systems that also harbor the most massive giant planets. On the other hand, the presence of an outer giant companion does not have a meaningful impact on the mass of inner small planets, now favoring hypotheses such as the ones from \citet{schlecker}, that says that giant planets are likely to destroy inner super-Earths; or \citet{bitsch}, who suggested that too massive cold Jupiters prevent the survival rate of inner systems.

Given that the positive correlation is present only for the least massive giants, we reaffirm mass dependence as a feasible explanation to the discrepancies in the literature.
The studies that find a strong correlation \citep{zhu2018, bryan2019, rosenthal2022} contain mostly lower mass giants, while the study from \citet{bonomo} appears to include more massive giant planets. This agrees with the proposed hypothesis and provides a possible explanation for the non-detection of the correlation. Further analysis on that point can be done considering the sensitivity of each survey, or utilizing larger datasets, in order to test the hypothesis more strictly.

We also emphasize the impact of these results on current surveys: surveys looking for super-Earth companions of giant planets, such as Gaia \citep{espinoza}, may not be as successful as hoped if their target stars are mostly hosts of super-Jupiters. This would also be a good test to the proposed hypothesis (i.e. the positive correlation between super-Earths and cold Jupiters holds only for the least massive giant planets).

Finally, if the presence of a super-Earth does depend on the mass of the giant companion, the absence of super-Earths in the Solar System could be explained by Jupiter belonging to the group of massive giant planets. However, Jupiter is below our defined projected mass threshold ($1.5 M_J$), and would fit into the group of Saturns (more likely to have super-Earths), making the Solar System an outlier. Still, the boundary was defined just by dividing the sample in two, so a better threshold can be determined by studying a larger sample.

\section*{Acknowledgements}
We thank the anonymous referee for their useful comments and suggestions that helped improve this paper.
This research has made use of data from the California Legacy Survey, so we are grateful to Lee Rosenthal, Benjamin Fulton, and all the researchers involved. G.D.M. acknowledges support from FONDECYT projects 1252141 and 11221206, and the ANID BASAL project FB210003. E.L. acknowledges support from FONDECYT project 11221206. E.L. is grateful to the support from from Pilar Forján and Jean-Marie Lefèvre.
We thank Cristóbal Petrovich, Claudia Aguilera, Michele de Leo, Cristóbal Zilleruelo, María Jesús Flores and Manuel Cavieres for their helpful comments and discussions about this research.

\appendix
\section{Calculation of the intrinsic number of planets}\label{appA}

In this appendix, we further explain the process of the detection completeness corrections we performed.

First, we define two-dimensional bins for the samples: $9$ bins evenly spaced in log distance between $1.5 \cdot 10^{-2}$ and $41.6$ au, and $9$ bins evenly spaced in log mass between $1.9$ and $9.8 \cdot 10^3 \, M_\oplus$. Then, we calculate the average detection completeness inside each bin. Finally, we divide the total number of cold Jupiters inside each bin by the corresponding average detection completeness.

\begin{figure*}[hbt!]
\label{fig:7}
    \centering
    {{\includegraphics[width=240pt]{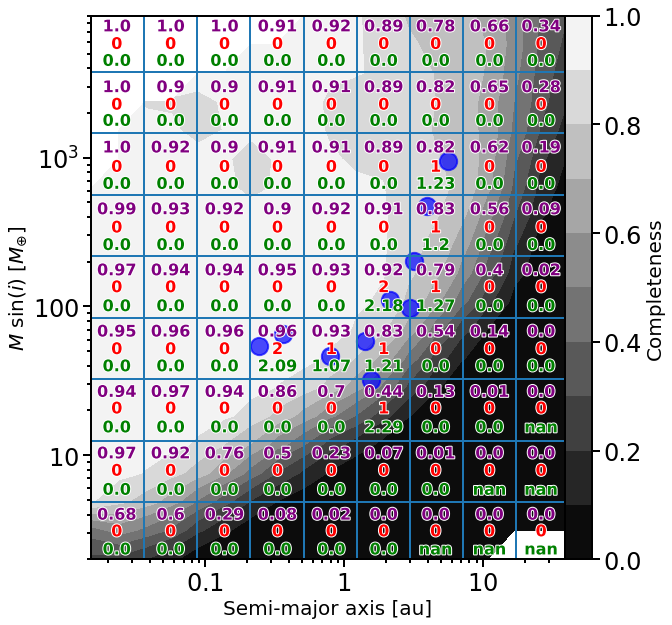} }}
    \qquad
    {{\includegraphics[width=240pt]{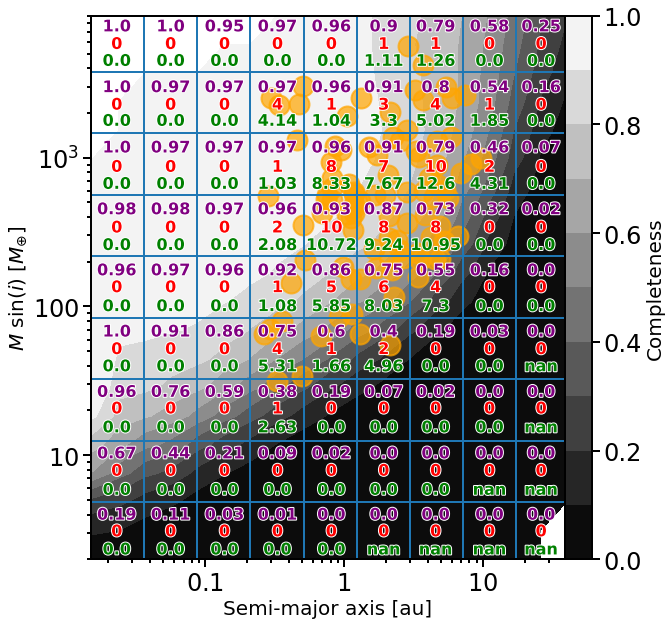} }}
    \caption{Average completeness (purple), number of planets before correction (red) and number of planets after correction (green) inside each bin (blue lines). Blue and orange circles represent cold Jupiters with and without inner super-Earth companions respectively. The contours on the left panel correspond to detection completeness of stars from the sample hosting at least one super-Earth. On the right panel, the contours correspond to detection completeness of stars that do not host super-Earths. Note that dividing the red number by the purple number gives an approximation of the green number.}
\end{figure*}

For instance, pay attention to the bin with the most massive cold Jupiter with super-Earth companions (i.e. the bin that contains the blue circle with the highest projected mass value) from the left panel of \hyperref[fig:7]{figure 7}. The number of cold Jupiters with inner super-Earth companions in that bin is just 1. The corrected number of cold Jupiters with super-Earth companions is the red number divided by the purple one: $\frac{1}{0.82} \approx 1.2$. We can replicate this for the same bin in the right panel: the corrected number of cold Jupiters without super-Earth companions is $\frac{10}{0.79} \approx 12.7$. Summing over semi-major axis for each projected mass bin, we get the debiased histogram shown in \hyperref[fig:4]{figure 4}.
This same process is applied to super-Earths with and without outer giant companions.

\end{document}